\RequirePackage{fix-cm}
\documentclass[smallextended,
  envcountsect,]{svjour3}       
\smartqed  
\usepackage{graphicx}
\usepackage[round]{natbib}
\usepackage[dvipsnames,table,xcdraw]{xcolor}
\usepackage{subcaption}
\usepackage{listings}
\usepackage{relsize}
\usepackage{xspace}
\usepackage{orcidlink}

\usepackage[scaled=0.8]{FiraMono}
\usepackage[T1]{fontenc}

\usepackage[colorlinks=true,linkcolor=MidnightBlue,citecolor=PineGreen,urlcolor=blue]{hyperref}

\usepackage{graphicx}
\usepackage{tabularx}
\usepackage{multirow}
\usepackage{pdflscape}
\usepackage{makecell}

\usepackage{booktabs}

\usepackage[most]{tcolorbox}
\newtcolorbox{intentionbox}[2][]{enhanced,
    sharp corners,
    colback=white,
    colbacktitle=white,
    coltitle=black,
    boxrule=1pt,
    left=2mm,
    right=2mm,
    bottom=2mm,
    top=2mm,
    boxed title style={colframe=white},
    attach boxed title to top left={yshift=-3mm, xshift=2mm},
    center title,
    title=#2,#1
}
\newtcolorbox{rqbox}[2][]{enhanced,
    sharp corners,
    colback=white,
    colbacktitle=white,
    coltitle=black,
    boxrule=1pt,
    left=5mm,
    right=5mm,
    bottom=2mm,
    top=4mm,
    boxed title style={colframe=white},
    attach boxed title to top center={yshift=-3mm},
    center title,
    title=#2,#1
}
\newtcolorbox{rqanswerbox}[2][]{enhanced,
    minipage boxed title*=-3cm,
    sharp corners,
    colback=white,
    colbacktitle=white,
    coltitle=black,
    boxrule=1pt,
    left=5mm,
    right=5mm,
    bottom=2mm,
    top=4mm,
    boxed title style={colframe=white},
    attach boxed title to top center={yshift=-3mm},
    center title,
    title=#2,#1
}
\usepackage{enumitem}

\usepackage{tikz}
\newcommand*{\myhash}{%
  \begin{tikzpicture}
    \pgfmathsetlengthmacro\myWidth{.8*width("=")}%
    \pgfmathsetlengthmacro\myHeight{height("H")}%
    \pgfmathsetlengthmacro\mySepY{.3333*\myWidth}%
    \pgfmathsetlengthmacro\mySideBearing{.1*\myWidth}%
    \def\myAngle{70}%
    \pgfmathsetlengthmacro\mySepX{\mySepY/sin(\myAngle)}%
    \pgfmathsetlengthmacro\mySlantX{\myHeight/tan(\myAngle)}%
    \draw[line cap=round]
      (0, {(\myHeight - \mySepY)/2}) -- ++(\myWidth, 0)
      (0, {(\myHeight + \mySepY)/2}) -- ++(\myWidth, 0)
      ({(\myWidth - \mySepX - \mySlantX)/2}, 0)
      -- ({(\myWidth - \mySepX + \mySlantX)/2}, \myHeight)
      ({(\myWidth + \mySepX - \mySlantX)/2}, 0)
      -- ({(\myWidth + \mySepX + \mySlantX)/2}, \myHeight)
    ;%
    \useasboundingbox
      (-\mySideBearing, 0)
      (\myWidth + \mySideBearing, \myHeight)
    ;%
  \end{tikzpicture}%
}

\newcommand{\testcube}{\emph{TestCube}~\includegraphics[height=8pt, trim=2pt 3pt 2pt 0]{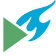}\xspace}

\definecolor{todo}{HTML}{6699ff}

\makeatletter
\newcommand{\hyperlinkTracer}[5]{
\ifx u#3\hyperlink{#1 #4}{(#2\myhash #4)}\else%
\ifx d#3\Hy@raisedlink{\hypertarget{#1 #4}{}}(#2\myhash #4#5)\else%
\fi\fi%
}%
\makeatother
\newcommand{\intention}[2][u]{%
  \hyperlinkTracer{intention}{I}{#1}{#2}{}%
}
\newcommand{\observation}[2][u]{%
  \hyperlinkTracer{observation}{O}{#1}{#2}{}%
}
\newcommand{\supobservation}[3][u]{%
  \hyperlinkTracer{observation}{O}{#1}{#2}{, Support: #3}%
}
\newcommand{\choice}[2][u]{%
  \hyperlinkTracer{choice}{C}{#1}{#2}{}%
}

\newcommand{\RQ}[1] {\textbf{RQ#1}}

\newcommand{\support}[1] {(Support: #1)}

\newcommand\circlenum[1]{\raisebox{1.2pt}{\textcircled{\hspace{0.35pt}\scriptsize{\raisebox{-.4pt}{#1}}}}}

\usepackage{cleveref}

\begin{document}
\title{Developer-Centric Test Amplification
}
\subtitle{The Interplay Between Automatic Generation\\and Human Exploration}


\author{Carolin Brandt         \and
        Andy Zaidman 
}


\institute{Carolin Brandt\, \orcidlink{0000-0001-7623-1970} \at
              Delft University of Technology \\
              \email{c.e.brandt@tudelft.nl}           
           \and
           Andy Zaidman\, \orcidlink{0000-0003-2413-3935} \at
              Delft University of Technology \\
              \email{a.e.zaidman@tudelft.nl}
}

\date{Received: date / Accepted: date}

\maketitle

\begin{abstract}
Automatically generating test cases for software has been an active research topic for many years.
While current tools can generate powerful regression or crash-reproducing test cases, these are often kept separately from the maintained test suite.
In this paper, we leverage the developer's familiarity with test cases amplified from existing, manually written developer tests.
Starting from issues reported by developers in previous studies, we investigate what aspects are important to design a developer-centric test amplification approach, that provides test cases that are taken over by developers into their test suite.
We conduct 16 semi-structured interviews with software developers supported by our prototypical designs of a developer-centric test amplification approach and a corresponding test exploration tool.
We extend the test amplification tool DSpot, generating test cases that are easier to understand.
Our IntelliJ plugin \testcube empowers developers to explore amplified test cases from their familiar environment.
From our interviews, we gather 52 observations that we summarize into 23 result categories and give two key recommendations on how future tool designers can make their tools better suited for developer-centric test amplification.

\end{abstract}

\iftrue

\section{Introduction}
\label{sec:intro}

Testing is an important~\citep{whittaker2012google}, but time-consuming activity in software projects~\citep{beller2015when,beller2015how,beller2019developer}.
Automatic test generation aims to alleviate this effort by reducing the time developers spend on writing test cases.
The software engineering community has created a plethora of powerful tools, that can automatically generate JUnit test cases for software projects written in Java.
For example, a widely known tool is EvoSuite~\citep{fraser2011evosuite}, which generates test cases from scratch using search-based algorithms.
It starts from a group of randomly generated test cases and optimizes them by mutating their code and combining them with each other.
This paper focuses on \emph{test amplification}, a technique that automatically generates new test cases by adapting existing, manually written test cases~\citep{danglot2019snowballing}.
The state-of-the-art test amplification tool DSpot~\citep{danglot2019automatic} mutates the setup phase of manually written test cases and generates new assertions to test previously untested scenarios.
For both EvoSuite and DSpot, studies have shown that the tools are effective in generating or extending test suites to reach a high structural coverage and mutation score~\citep{danglot2019automatic,fraser2011evosuite,rojas2015automated,serra2019on-the-effectiveness}.

Automatic test generation is, for example, used to detect regressions~\citep{robinson2011scaling}, reproduce crashes~\citep{derakhshanfar2020good,derakhshanfar2020botsing}, uncover undertested scenarios~\citep{stamp2019use-cases} and generate test data~\citep{haq2021automatic}. 
For these use cases, it is often sufficient to keep the generated test cases separate from the manually written and maintained test suite~\citep{stamp2019use-cases,nassif2021generating}.
This separation is reinforced by several hard-to-solve challenges that limit the understandability of the automatically generated tests, such as their readability~\citep{daka2015modeling,grano2018an-empirical} or generating meaningful names~\citep{zhang2016towards,daka2017generating}, or documentation~\citep{roy2020deeptc,panichella2016the-impact,bihel2018adapting}.

The amplified test cases created by DSpot are closely based on manually written ones.
This opens up the chance to generate test cases that are easier to understand by developers, as they are likely familiar with the original test case, which the amplified test case is based on.
In this paper, we want to leverage this aspect and take a look at generating amplified test cases that developers can take over into the manually maintained test suite as if they would have written them themselves.
To describe this kind of test generation we use the term \emph{developer-centric}, as the developer accepting the test case is central for this kind of test generation.

Generated test cases that are accepted by developers and are part of the maintained test suite also fulfill several further typical uses for developer tests.
For example, as a form of executable documentation~\citep{hoffman2003api-documentation,beck2003test-driven,kochhar2019practitioners}, or to locate the fault that causes a failing test by understanding the test in question~\citep{panichella2016the-impact}.

To provide amplified test cases that developers take over into their test suite, the interaction of the developer with the test amplification tool in which they review the proposed amplified test cases is critical.
In past projects, users of DSpot reported that the tool was complex to configure and they had to wait long for the tool to finish and for them to see results~\citep{stamp2019use-cases}.
There is little support that guides developers through the list of generated test cases so they can effectively judge whether to keep or discard a newly amplified test case.
To address these issues and realize developer-centric test amplification, we embed the test amplification tool in a so-called \emph{test exploration tool}:
\begin{quote}
  A \emph{test exploration tool} forms the interaction layer between the developer and the test amplification tool. It lets the developer start the test amplification tool, and later explore and inspect the different amplified test cases.
\end{quote}

To illustrate how a developer would use a developer-centric test amplification approach with a test exploration tool, we introduce an exemplary use case, which is also illustrated in \Cref{fig:overview}:
\begin{quotation}
Hannah, a software developer, wants to expand the test suite of the industrial project she is working on to cover more functionality and give her confidence that they are not breaking important behavior when changing something.
As her management is constantly asking for new features, she is pressed on time and decides to use an automatic test amplification tool to improve her project's test suite.
From her integrated development environment (IDE), she starts the test amplification.
The tool generates several new test cases for her and notifies her that it is finished.
Hannah inspects the test cases one by one directly from the test exploration tool integrated into her IDE.
The exploration tool shows her the code of each new test case and where in the production code new instructions are covered.
Hannah browses through the new test cases and if she is happy with any test case, she adds it to the test suite with one click of a button.
After exploring all proposed test cases, she commits her changes and can lean back with the confidence of a better tested system.
\end{quotation}

\begin{figure*}[htb]
		\centering
		\includegraphics[width=\textwidth, clip,
    ]
    {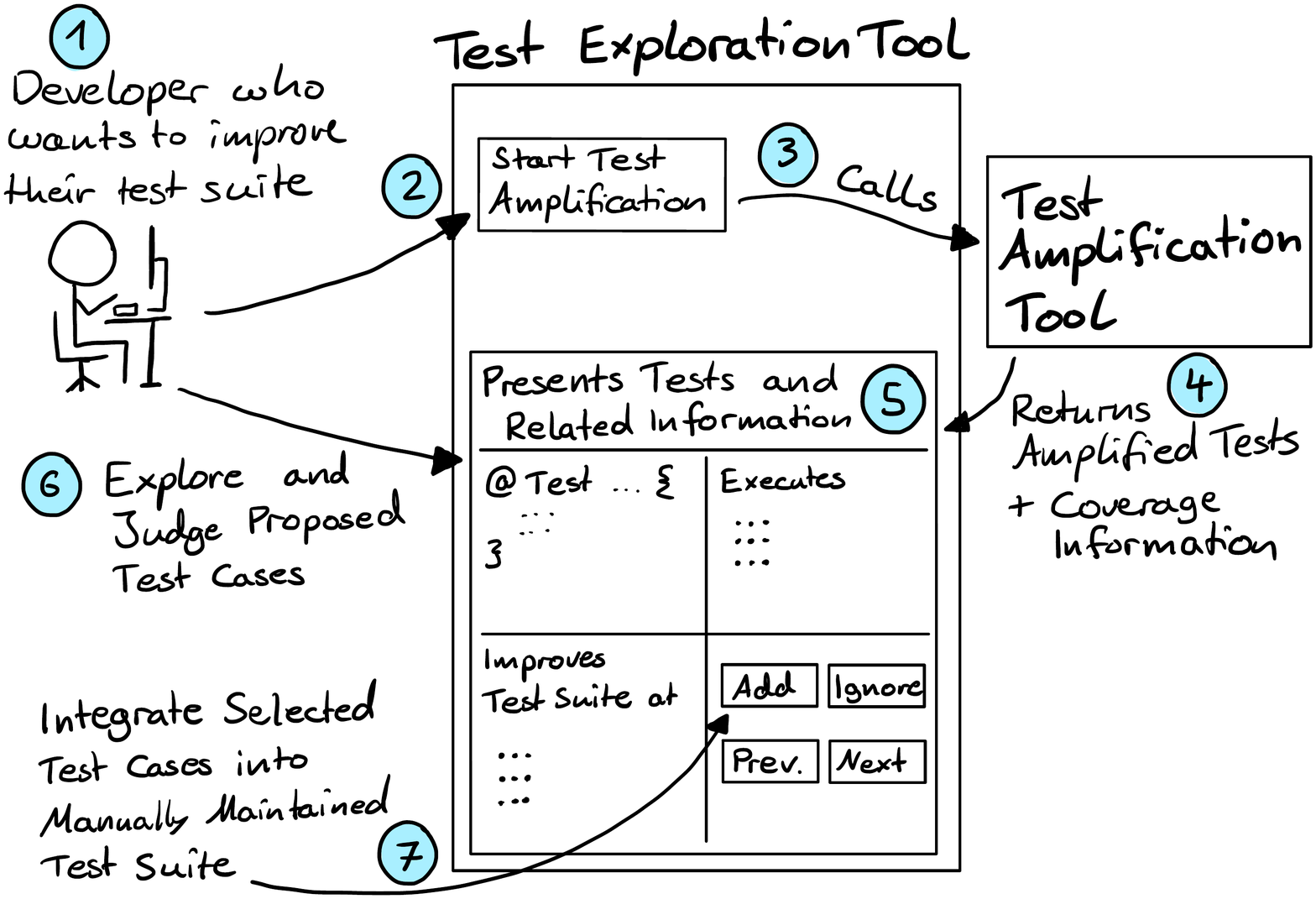}
		\caption{Overview of test amplification with the help of a test exploration tool.}
		\label{fig:overview}
\end{figure*}

In this paper, we investigate how we should design a developer-centric test amplification approach to be successful with developers.
As automatic test amplification is not widely used and to prevent re-studying the already known issues in current tools, we develop prototypes of a developer-centric test amplification approach and a corresponding test exploration tool.
To motivate the design choices we take for our prototypes, we derive four design intentions from the test generation literature and the use case we propose for developer-centric test amplification.
Based on these intentions we revise the test amplification process of DSpot to generate shorter, easier to understand test cases.
Our corresponding IntelliJ plugin \testcube lets the developer generate and explore test cases right from their integrated development environment (IDE).

We conduct a qualitative study to explore which aspects of our prototypes are successful in supporting developer-centric test amplification and uncover further aspects that should be addressed to realize it.

In previous studies, developers using EvoSuite were ``concerned about the readability of generated unit tests, the generated input data, and the generated assertions''~\citep{almasi2017an-industrial}, while DSpot users found it difficult to understand the generated test cases~\citep{stamp2019use-cases}.
Stemming from these observations, we investigate in our first research question what developers find important in the code and behavior of an amplified test case.
The answers to this question give guidance on what factors in amplified test cases are relevant for developers to include the test cases in their maintained test suite.

\begin{intentionbox}{\textbf{Research Question 1}}
  What are the key factors to make amplified test cases suited for developer-centric test amplification?
\end{intentionbox}

In our second research question, we explore how test exploration tools should be designed to support developer-centric test amplification.

\begin{intentionbox}{\textbf{Research Question 2}}
  What are the key factors to make test exploration tools suited for developer-centric test amplification?
\end{intentionbox}

A powerful capability of such test exploration tools is to provide the developer with information beyond the test code itself.
We already know from test review that developers are interested in understanding the code under test, as well as knowing the code coverage of test cases~\citep{spadini2018when}.
To deepen our insights into what information test exploration tools should make accessible to developers, we pose our third research question.

\begin{intentionbox}{\textbf{Research Question 3}}
  What information do developers seek while exploring amplified test cases?
\end{intentionbox}

Creating a great tool alone is not enough for developers to appreciate using it.
We also need to convey the value our tool brings to them.
Therefore our fourth research question asks what value developers can gain from using developer-centric test amplification.
With the answers to this question, future tool creators know which benefits they can focus on when they seek to convince users to start or keep using their tool.

\begin{intentionbox}{\textbf{Research Question 4}}
  What value does developer-centric test amplification bring to developers?
\end{intentionbox}

To answer these research questions, we conduct semi-structured interviews with 16 software developers from varied backgrounds.
The participants tried out our prototypes and provided us with rich insights on their impressions of our prototypes and how we could improve them to better fit their needs.
We group 52 recurring observations from the interviews in 23 result categories for our four research questions.
During the discussion of these results, we identify two key recommendations on how we should design future developer-centric test amplification tools.

With this paper, we are taking a step towards developer-centric test amplification.
In short, we contribute:
\begin{itemize}
    \item two recommendations on how to design developer-centric test amplification tools
    \item a structured overview of the key factors to make amplified tests as well as test exploration tools suited for developer-centric test amplification
    \item a refined, developer-centric test amplification approach, based on the DSpot test amplification
    \item a developer-oriented test exploration plugin for the IntelliJ IDE
\end{itemize}

\section{Creating Developer-Centric Test Amplification}
\label{sec:intentions}

In this paper, we aim to investigate the key aspects that make amplified test cases and test exploration tools suited for developer-centric test amplification by conducting semi-structured interviews with software developers.
To illustrate the concept of test amplification to our participants and receive rich and concrete input, we want to let them try out a test amplification tool during the interview.
A state-of-the-art test amplification tool is DSpot~\citep{danglot2019automatic}, which was developed and evaluated during the European H2020 STAMP project.
We adapt DSpot's amplification process based on the feedback from the reports of the European project~\citep{stamp2019use-cases} and the requirements posed by our use case of developer-centric test amplification.
To facilitate the interaction of the developer with the test amplification we also design a prototype of a test exploration tool.

In this section we discuss the inspirations leading to our design of both prototypes, which is described in \Cref{sec:design}. The goal of this section is to clarify our reasoning behind the design choices we took and connect them to the existing literature and user reports about DSpot.
We formulate four design intentions and present their connection to our design choices in \Cref{tab:intentions-and-decisions}.

A central part of the load on developers comes from them having to understand the test cases and judge whether they check intended behavior.
According to Meszaros, obscure tests that are difficult to understand at a glance are an anti-pattern, as it makes tests harder to maintain and potential bugs in the test code more difficult to detect~\citep{meszaros2007xunit}.
As automatically generating human-readable code is a hard problem to solve, readability and understandability of generated test cases are recurring topics in developer's feedback:
developers from the STAMP project stated about DSpot that it was hard to interpret the tool's output and the ``resulting tests were difficult to understand for a human developer''~\citep{stamp2019use-cases}.
In some cases, the developers found the generated tests useful, but so hard to read that they wrote a corresponding test case themselves.
Nevertheless, they were glad that DSpot pointed to real bugs and supported them in testing exceptional behavior in systems where only the optimal behavior was tested before.
Several previous works in test generation were concerned with making the generated tests more understandable for developers~\citep{panichella2016the-impact,daka2017generating,roy2020deeptc,palomba2016automatic,rojas2015automated}.
This clearly shows that we should also take understandability into account while designing our prototypes.
Therefore, our first design intention is to generate test cases that are understandable for developers.

\begin{intentionbox}{\textbf{Intention 1 \intention[d]{1}}}
  Generate test cases that are understandable for developers
\end{intentionbox}

From literature and our own experiences, we understand that the users of current test amplification tools face obstacles that lead them to abandon automatic test amplification.
During the STAMP project~\citep{stamp2019use-cases}, various industrial partners noted that DSpot takes very long to generate test cases.
The configuration is overly ``complex because of the multitude of possible parameter values'' which require experience to tweak correctly.
The high effort required by users was reported for other test generation tools, too.
Previous studies of EvoSuite pointed out the high load on developers to inspect generated test suites and to decide if assertions in test cases are correct~\citep{fraser2015does}.
They also spend a lot of time analyzing generated test cases to decide whether to improve or discard them~\citep{fraser2015does} as generated test cases tend to be less readable than manually written ones~\citep{grano2018an-empirical}.
That is why another intention leading the design of our prototypes is to decrease the load on the developers while they use test amplification.



\begin{intentionbox}{\textbf{Intention 2 \intention[d]{2}}}
  Easy interaction to decrease the load on the developer
\end{intentionbox}

Another intention leading our design, is that the amplification process should be fast enough so that developers can start it and receive new test cases in the same session.
This means, for example, that they do not have to wait for an external build process to finish.
We conjecture that this makes it easier for them to understand the results and the value of the test amplification as they can include it directly while they work on improving their test suite.

\begin{intentionbox}{\textbf{Intention 3 \intention[d]{3}}}
  Fast enough for direct interaction
\end{intentionbox}

Lastly, it is our intention that the developers can grasp the impact an amplified test case.
They should see the test case as a useful addition when taken over into their test suite.
Impact in this case could refer, for example, to the coverage, code quality, test code size or test suite runtime.
We assume that understanding the impact is a pre-requisite to deciding whether the test is useful or not.
The test exploration tool should make the impact and the quality of the amplified test cases clear so that the developers see the value that the automatic test amplification brings them.

\begin{intentionbox}{\textbf{Intention 4 \intention[d]{4}}}
  Impact is clear to developers and they find the tests useful
\end{intentionbox}

Both \intention{2} and \intention{4} can not be addressed by modifying the test amplification of DSpot itself.
Rather this shows the need for a layer in between the test amplification tool and the developer that facilitates their interaction.
This role is taken by the test exploration tool.

\section{Bringing Test Amplification to the Developer (IDE)}
\label{sec:design}

Based on the intentions we defined in \Cref{sec:intentions}, we develop prototypes for both a developer-centric test amplification tool as well as a test exploration tool.
We will use these prototypes during our interviews to illustrate a possible version of developer-centric test amplification.
In the following, we present our design and explain how our choices are motivated by the intentions we set.
\Cref{tab:intentions-and-decisions} gives an overview of these choices, which we mark throughout the text with \choice[d]{n}.

\newcommand{\rotatedcell}[1] {\rotatebox{90}{\parbox[b]{4.3cm}{#1}}} 
\newcommand{\choicecell}[1] {\rotatebox{90}{\parbox[b]{1cm}{\raggedleft \choice{#1}}}}

\begin{table}[]
\setlength{\tabcolsep}{3pt}
\centering
\begin{tabular}{l | ccccccccc | cccccc |}
\multicolumn{1}{l}{} & \multicolumn{15}{c}{Design Choices}
\\
\multicolumn{1}{l}{} & \multicolumn{9}{c}{Test Generation}  & \multicolumn{6}{c}{\makecell{Test Exploration}}
\\ \cmidrule(lr){2-10} \cmidrule(lr){11-16}
& \choicecell{1} & \choicecell{2} & \choicecell{3} & \choicecell{4} & \choicecell{5} & \choicecell{6} & \choicecell{7} & \choicecell{8} & \choicecell{9} & \choicecell{10} & \choicecell{11} & \choicecell{12} & \choicecell{13} & \choicecell{14} & \choicecell{15}
\\
\makecell{Intention} & \rotatedcell{Amplification} & \rotatedcell{Remove calls inside old assertions} & \rotatedcell{One input mutation} & \rotatedcell{Explanatory comments} & \rotatedcell{One assertion generated} & \rotatedcell{Assertion matching input} & \rotatedcell{Instruction coverage} & \rotatedcell{Look for additonally covered inst.} & \rotatedcell{Report coverage} & \rotatedcell{IDE plugin} & \rotatedcell{Start with run} & \rotatedcell{Background task} & \rotatedcell{Default configuration} & \rotatedcell{Coverage information text} & \rotatedcell{Coverage editor}
\\ \midrule
\makecell{Generate test cases that\\are understandable for\\developers \intention{1}} & x & x & x & x & x & & & x & & & & & & & 
\\ \midrule
\makecell{Easy interaction to\\decrease the load on\\the developer \intention{2}} & x & & & & & & & & & x & x & x & x & & 
\\ \cmidrule{1-16}
\makecell{Fast enough for direct\\interaction \intention{3}} & & & & & & & x & & & & & & & & 
\\ \cmidrule{1-16}
\makecell{Impact is clear to\\developers and they find\\the tests useful \intention{4}} & & & & & & x & & & x & & & & & x & x
\\ \midrule
\end{tabular}
\setlength{\tabcolsep}{6pt}
\caption{The relation between our design intentions (I\protect\myhash 1-4) and the design choices we take for our developer-centric test amplification and exploration prototypes (C\protect\myhash 1-15).}
\label{tab:intentions-and-decisions}
\end{table}

This section starts with a more detailed definition of test amplification and clarifies why we choose to base our developer-centric test generation on this technique. 
We describe how we adapt DSpot to generate test cases that are better suited to be read by developers.
Further, we present our test exploration tool, the IntelliJ Plugin \testcube, which enables developers to easily use our developer-centric test amplification with minimal configuration, right from their familiar development environment.

\paragraph{Introduction to Test Amplification}
Test amplification is a term for test generation techniques that take manually written test cases as their primary input.
Danglot et al. conducted a literature study to map this emerging field and defined test amplification as follows:
\begin{quote}
  Test amplification consists of exploiting the knowledge of a large number of test cases, in which developers embed meaningful input data and expected properties in the form of oracles, in order to enhance these manually written tests with respect to an engineering goal (e.g., im- prove coverage of changes or increase the accuracy of fault localization).~\citep{danglot2019snowballing}
\end{quote}
For our prototype design we choose test amplification to generate the test cases.
We exploit the existing test cases, as well as the code under test to create additional test cases that improve the instruction coverage of a test suite.

We base our approach on test amplification \choice[d]{1}, because we expect that for a developer already familiar with the test suite it will be easier to understand a variation of an existing test case than a completely new one.
In addition, most software projects that are looking to improve their testing already have at least a rudimentary test suite.

\subsection{Developer-Centric Amplification with DSpot}

During our interviews, we want to showcase a possible version of developer-centric amplified test cases to software developers.
We adapt Danglot et al.'s tool DSpot~\citep{danglot2019automatic}, addressing the issues which were already reported by developers.
\Cref{fig:test-amp-explanation} gives an overview of our revised test amplification approach.
Starting with the original test case from the existing test suite, we remove all existing assertions, modify the objects and values in the setup phase of the test case, add new assertions based on the changed behavior, and select test cases that cover additional instructions in the code under test.

Our design and implementation is strongly based on Danglot et al.~\citep{danglot2019automatic} and DSpot version 3.1.0\footnote{\url{https://github.com/STAMP-project/dspot/releases/tag/dspot-3.1.0}}.
We created a fork of their repository\footnote{\url{https://github.com/TestShiftProject/dspot/releases/tag/v3.2.0-dev-friendly}} and contributed our changes back to DSpot through an accepted pull request\footnote{\url{https://github.com/STAMP-project/dspot/pull/993}}.
In the following, we describe for each step the behavior of the original amplification as well as the changes we made to generate more understandable test cases \intention{1} and convey their value to developers more easily \intention{4}.
We illustrate our explanations with a running example in \Cref{lst:remove-assertions,,lst:mutate-input,,lst:generate-assertion,,lst:select-test-case}.

\begin{figure*}[htb!]
		\centering
		\includegraphics[width=\textwidth, clip,
    trim={0.3cm 28cm 15.5cm 0cm},
    page=1
    ]
    {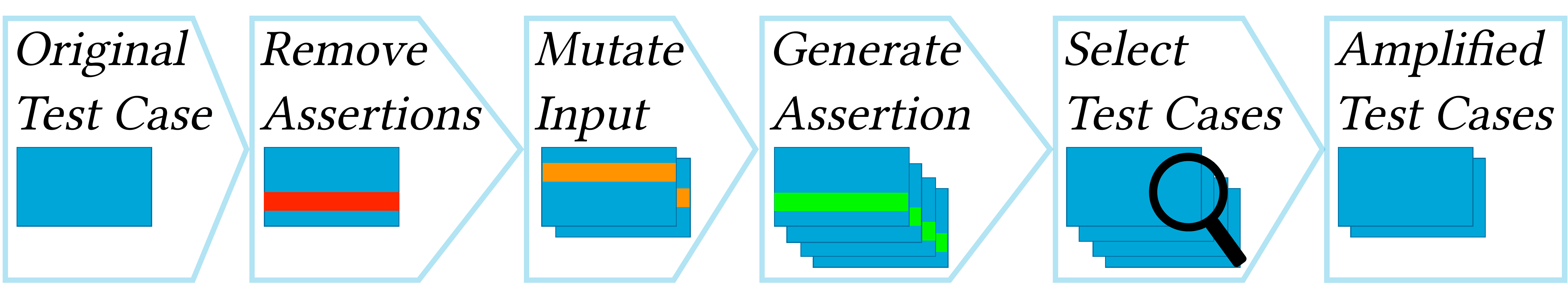}
		\caption{Overview of our automatic, developer-centric amplification process within DSpot.}
		\label{fig:test-amp-explanation}
\end{figure*}

\subsubsection{Remove Assertions}
At the start of the amplification process, DSpot removes all assertions in the original test case, as they will likely no longer match the new amplified test case.
All method calls within the assertions are preserved because they might have side effects that influence the rest of the method calls in the test case.

However, these method calls tend to be confusing outside of the context of the assertion.
As the behavior of the test cases is changed through the amplification anyways, we decide to also remove method calls within assertions \choice[d]{2}.
\Cref{lst:remove-assertions} shows the first amplification step for our example, where the two assertions of a test case are removed completely.

\lstset{
  language=Java,
  basicstyle=\ttfamily\footnotesize,
  morekeywords={Examples, Example, Log, Keywords, Category, Chunk},
  postbreak=\mbox{\textcolor{cyan}{$\hookrightarrow$}\space},
	escapeinside=**,
  breaklines=true,
  showspaces=false,
  showstringspaces=false,
  commentstyle = \color{CadetBlue},
  keywordstyle = \color{blue},
  stringstyle = \color{Plum},
  rulecolor = \color{black},
  extendedchars=true,
  texcl=false,
  aboveskip=\baselineskip,
  belowskip=0pt,
}
\begin{figure}[tbp]
	\centering
  \begin{lstlisting}
public class AttributeTest {
    @Test
    public void html() {
        Attribute attr = new Attribute("key", "value &");
  \end{lstlisting}
  \vspace{-\baselineskip}
  \begin{lstlisting}[backgroundcolor=\color{Red!40}]
-       assertEquals("key=\"value &amp;\"", attr.html());
-       assertEquals(attr.html(), attr.toString());
  \end{lstlisting}
  \vspace{-\baselineskip}
  \begin{lstlisting}
    }
}
  \end{lstlisting}
\vspace{-0.3cm}
	\caption{Example amplification: Remove all existing assertions from the original test case.}
	\label{lst:remove-assertions}
\vspace{-0.3cm}
\end{figure}

\subsubsection{Mutate Input}
DSpot uses a variety of mutations to explore the input space of a test case.
Literals like integers, booleans, and strings are slightly modified or replaced by completely random values.
On existing objects, the input amplification removes, duplicates, or adds new method calls.
It can also create new objects or literals that are then used as parameters for mutated method calls.

Our developer-centric amplification leverages the powerful input mutations of DSpot.
However, from Grano et al. we know that the more complex a test case is, the harder it is to understand for a developer~\citep{grano2020pizza}.
To make the generated test case easier to understand for the developer, we focus on one input modification at a time \choice[d]{3} and add an explanatory comment to every mutation \choice[d]{4}.
We make use of all available mutation operations in DSpot 3.1.0.
In \Cref{lst:mutate-input} one of the string parameters in the constructor for the object \texttt{attr} is replaced with a new string that contains the special \emph{non-breaking space} character.
We also add a comment that details which value was changed to which new value to help the developer spot the change easily.

\begin{figure}[tbp]
	\centering
  \begin{lstlisting}
public class AttributeTest {
    @Test
    public void html() {
  \end{lstlisting}
  \vspace{-\baselineskip}
  \begin{lstlisting}[backgroundcolor=\color{Red!40}]
-       Attribute attr = new Attribute("key", "value &");
  \end{lstlisting}
  \vspace{-\baselineskip}
  \begin{lstlisting}[backgroundcolor=\color{Green!40}]
+       // FastLiteralAmplifier: change string from 'value &' to 'Hello\nthere*{\color{CadetBlue}$\overline{\underline{\text{NBSP}}}$}*'
+       Attribute attr = new Attribute("key", "Hello\nthere*{\color{Plum}$\overline{\underline{\text{NBSP}}}$}*");
  \end{lstlisting}
  \vspace{-\baselineskip}
  \begin{lstlisting}
    }
}
  \end{lstlisting}
\vspace{-0.3cm}
	\caption{Example amplification: Mutate string parameter in the constructor of the object under test.}
	\label{lst:mutate-input}
\vspace{-0.3cm}
\end{figure}

\subsubsection{Generate Assertion}
Generating new assertions is one of the central features of DSpot.
The tool instruments the test case to observe the state of the objects under test after the setup phase.
Then it generates assertions comparing the return value of every method call on the objects under test with the observed value.
While adding all generated assertions leads to a more powerful test case with respect to structural coverage, it also makes the test case hard to understand and unclear which of the added assertions improve these metrics.
To minimize the generated test cases, DSpot provides a \texttt{prettifier} stage.
It removes the assertions one by one, reruns the metric calculation, and adds the assertion back if the score decreased.
Unfortunately, the stage multiplies the already long runtime of DSpot.

To generate shorter, more understandable test cases \intention{1}, we opt to only add one assertion to each test case \choice[d]{5}.
While this at first generates more test cases, the ones with assertions that do not improve the final selection metric are excluded in the following step.
To produce test cases that developers find useful \intention{4}, the generated assertion should assert a behavior that changed through the previously mutated input.
To achieve this, we compare all assertion candidates before and after the mutation and only include an assertion if the value it asserts changed through the mutation \choice[d]{6}.

As shown in \Cref{lst:generate-assertion}, the assertion generated for our example checks the return value of \texttt{attr.toString()}, which shows the changed input \verb!"Hello\\nthere!\\$\overline{\underline{\text{NBSP}}}$\verb!"!.

\begin{figure}[tbp]
	\centering
  \begin{lstlisting}
public class AttributeTest {
    @Test
    public void html() {
        // FastLiteralAmplifier: change string from 'value &' to 'Hello\nthere*{\color{CadetBlue}$\overline{\underline{\text{NBSP}}}$}*'
        Attribute attr = new Attribute("key", "Hello\nthere*{\color{Plum}$\overline{\underline{\text{NBSP}}}$}*");
  \end{lstlisting}
  \vspace{-\baselineskip}
  \begin{lstlisting}[backgroundcolor=\color{Green!40}]
+       Assertions.assertEquals("key=\"Hello\nthere&nbsp;\"", ((Attribute) (attr)).toString());
  \end{lstlisting}
  \vspace{-\baselineskip}
  \begin{lstlisting}
    }
}
  \end{lstlisting}
\vspace{-0.3cm}
	\caption{Example amplification: Generate an assertion which checks a behavior changed by mutating the input.}
	\label{lst:generate-assertion}
\vspace{-0.3cm}
\end{figure}

\subsubsection{Select Test Cases}
\begin{figure}[tbp]
	\centering
  \begin{lstlisting}[backgroundcolor=\color{Cerulean!40}]
Amplified test case 'html_literalMutationString19_assSep92'
This test case improves the coverage in these classes/methods/lines:
org.jsoup.nodes.Entities:
escape
L. 197 +3 instr.
L. 198 +5 instr. 
  \end{lstlisting}
\vspace{-0.3cm}
	\caption{Example amplification: Information about the coverage improvement of the amplified test case.}
	\label{lst:select-test-case}
\vspace{-0.3cm}
\end{figure}

After generating a broad range of test cases through mutating input values and generating assertions, DSpot selects which test cases to keep.
Depending on the configuration, DSpot selects test cases that improve instruction coverage, improve mutation score, or cover the changes in a specific commit.

As determining the mutation score is computationally expensive, it is currently not a feasible option if the test generation should run on the developer's local computer and we want to enable direct interaction with as little wait time as possible \intention{3}.
Therefore, we select test cases based on instruction coverage \choice[d]{7}.

DSpot originally keeps all generated test cases that by themselves cover more lines than the original test case they are based on.
However, for a developer, it is not important that the coverage of one test case is high.
Rather, a new test case should contribute \emph{additional coverage} to the test suite.
To determine this, we measure the instruction coverage of the original test suite on a fine-grained, line-by-line basis.
For each generated test case, we check whether it covers \emph{additional} instructions on any line \choice[d]{8}.
If that is the case, we keep the test case, if not, we discard it.
The combination of this fine-grained coverage comparison together with the small number of additions we make to the original test case \choice{3} \choice{5} enables us to generate smaller test cases \intention{2} compared to DSpot.
Furthermore, these test cases have a local and therefore easier to understand impact on the coverage of the test suite \intention{4}.

To communicate to the developer which additional instructions are covered, our developer-centric amplification reports for each test case in which lines in the production code additional instructions are covered \choice[d]{9}.
\Cref{lst:select-test-case} shows a pretty-printed version of the additional information we provide.
The amplified test case in our example covers 8 more instructions over two lines in the \texttt{escape} method of the \texttt{Entities} class.

\subsection{Test Exploration Plugin \protect\testcube}
\label{sec:plugin}

\begin{figure*}[htb]
		\centering
		\includegraphics[width=\textwidth, clip,]
    {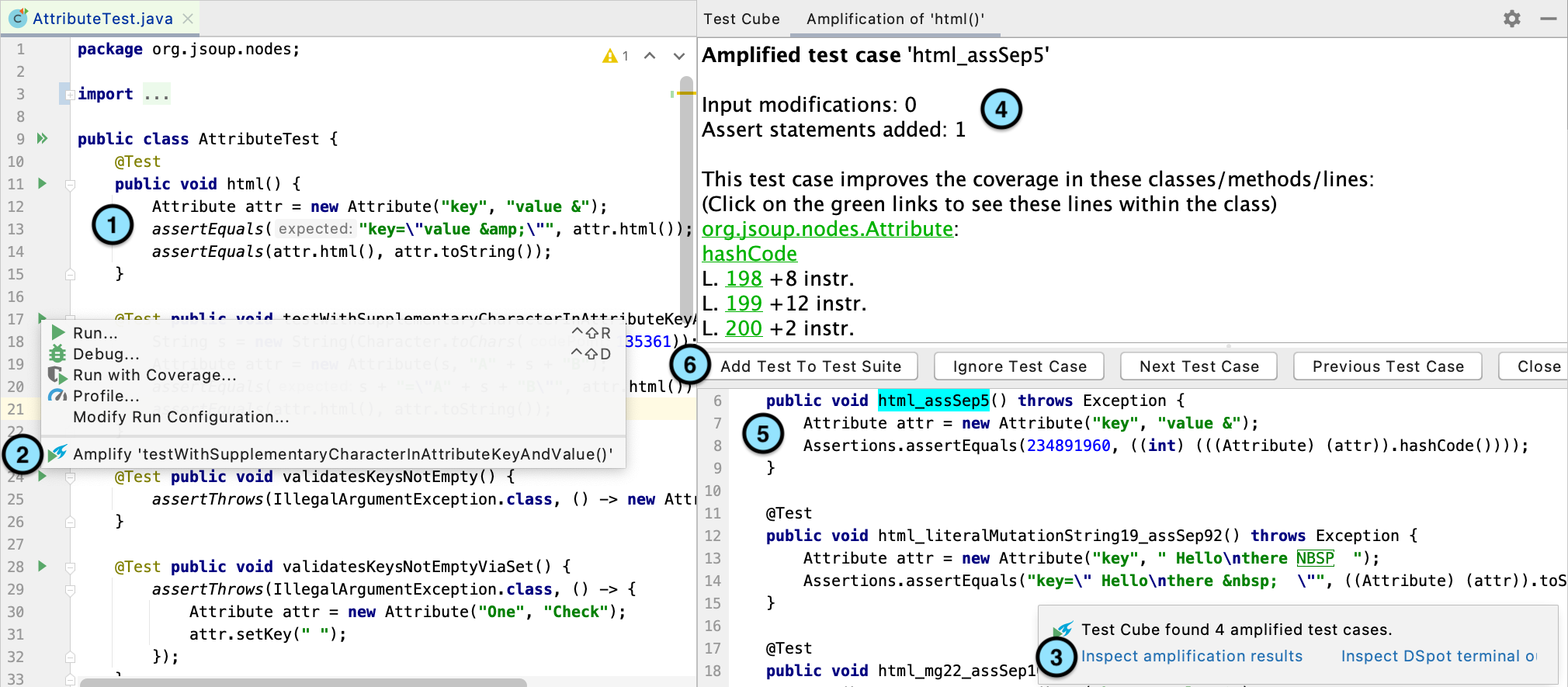}
		\caption{Overview of the interaction with the \emph{TestCube} plugin.}
		\label{fig:test-cube-overview}
\end{figure*}


For successful developer-centric test amplification, we conjecture that the second important step to support developers with amplification test cases is \emph{exploration}. 
In this section, we describe the design of our prototype of a developer-centric test exploration tool: \testcube.
To make our new, powerful test amplification easily accessible to developers \intention{2}, we develop \testcube as a plugin to the IntelliJ IDE \choice{10}.
A lot of previous research points out the importance of integrating tools into existing development environments.
It reduces time and focus lost by switching tools~\citep{liu2020exploring} and enables developers to inspect test cases and related code~\citep{spadini2018when}.
\testcube is open source\footnote{\url{https://github.com/TestShiftProject/test-cube}} and available to download on the JetBrains Marketplace\footnote{\url{https://plugins.jetbrains.com/plugin/14678-test-cube}}.
The screenshot in \Cref{fig:test-cube-overview} illustrates the user interface of \testcube and how a developer would interact with it.
In the following, we present how developers can use \testcube to amplify tests right from their editor, inspect the generated test cases and easily integrate them into their code base.

\subsubsection{Starting the Amplification}
After installing \testcube, the developer can start the amplification in the same way as she would execute a JUnit test \choice[d]{11}.
She picks an original test case to be the input for the amplification process (\circlenum{1} in \Cref{fig:test-cube-overview}).
Then she can click on the green arrow next to the  test, and select the new option `Amplify' (\circlenum{2} in \Cref{fig:test-cube-overview}).
After she selects this option, \testcube starts amplifying the test in a background task \choice[d]{12}.

For other test generation tools, research has shown that even though tuning parameters to the specific project increases performance, using the default settings already produces good results~\citep{arcuri2013parameter}.
To take the configuration burden off the developer as much as possible \intention{2} during our interviews and to evaluate how well our new approach performs with its default configuration, we choose to set default values for the vast number of DSpot parameters \choice[d]{13}.

\subsubsection{Result Inspection}
When the amplification finishes, \testcube notifies the developer with a pop-up from the built-in notification system (\circlenum{3} in \Cref{fig:test-cube-overview}).
The developer can then choose to inspect the test cases or, in case of reported errors, the terminal output of DSpot.

We present the results of the test amplification within IntelliJ, but in a tool window separated from the code.
The tool window is located on the right, next to the editor with the original test case selected by the developer.
It has various components:
\begin{itemize}
	\item At the top of the tool window we present information about the currently selected amplified test case (\circlenum{4} in \Cref{fig:test-cube-overview}):
Which modifications were applied to it and which additional instructions are covered \choice[d]{14}.
	\item Next we present \emph{navigation buttons} to the developer (\circlenum{6} in \Cref{fig:test-cube-overview}).
With these buttons, the developer cycles through the proposed test cases, can add the current test to the test suite, ignore the current test or close the amplification results all together.
\item Below the navigation buttons we present the amplified test cases in a fully functioning editor, as shown in (\circlenum{5} in \Cref{fig:test-cube-overview}).
The developer can edit the test case in their familiar environment and use code navigation to inspect called methods.
	\item The developer can click on the additionally covered lines in the \emph{test case information} to open up the \emph{coverage inspection editor} on the bottom of the tool window.
The editor shows the class where coverage was improved and highlights all additionally covered lines in green \choice[d]{15}.
Showing the added coverage in the context of the code under test should help the developer judge the value of the generated test case \intention{4}.
\end{itemize}

\section{Study Design}
\label{sec:evaluation}
The goal of this paper is to explore which aspects are important to create a successful developer-centric test amplification approach.
To this end, we invited 16 software developers to try out our prototype \testcube on an example project.
We observed their interaction with the tool and interviewed them on their experience and opinions on how an ideal test generation tool should behave.
We report our observations split along four sub-questions:
With the interviews we investigate what makes amplified tests (\RQ{1}) and exploration tools (\RQ{2}) suited for developer-centric test amplification, what information the developers are seeking while investigating the test cases (\RQ{3}) and what value developers see in test amplification (\RQ{4}).
In the following we describe the design of our interview study:
How we recruit participants and ask for their consent, our technical setup, the flow of one interview as well as our data collection and analysis process.

\begin{rqbox}{\textbf{Research Questions}}
\begin{itemize}[leftmargin=1cm]
  \item [\RQ{1:}] What are the key factors to make amplified test cases suited for developer-centric test amplification?
  \item [\RQ{2:}] What are the key factors to make test exploration tools suited for developer-centric test amplification?
  \item [\RQ{3:}] What information do developers seek while exploring amplified test cases?
  \item [\RQ{4:}] What value does developer-centric test amplification bring to developers?
\end{itemize}
\end{rqbox}

\subsection{Preparation}
We used convenience sampling to recruit participants for our interviews.
We posted about it on Twitter\footnote{\url{https://twitter.com/laci_noire/status/1328334375537299461}, showed to 9.527 users, 406 interactions} and wrote to existing industry contacts.
In addition, we contacted participants of a previous survey about motivation to write test cases who indicated to be open for a follow-up interview.

As we are conducting a study with human participants, we followed the guidelines of the TU Delft’s Human Research Ethics Council\footnote{\url{https://www.tudelft.nl/over-tu-delft/strategie/integriteitsbeleid/human-research-ethics}, last visited March 1st, 2021} and submitted our study design to them for review.
Before each interview, we explained to the interviewees how we will process their data and asked for consent on participating in the interview, recording the session for later analysis and publishing the anonymized results in an online research repository.
One participant wished to not be quoted and the corresponding results not to be published in a research repository, therefore our online appendix~\citep{brandt2021amplification-interview-replication} excludes the data collected from that interview.

To showcase \testcube to our participants, we selected the open-source HTML parser \emph{jsoup}\footnote{\url{https://github.com/jhy/jsoup}}.
Jsoup is a mid-sized Java project (35K lines of code), which is built with Maven, tested with JUnit 5 and was part of Danglot et al.'s evaluation of DSpot~\citep{danglot2019automatic}.
We chose it because we expected HTML to be a relatively simple and widely understood application domain, which would require less time to explain to our participants during the interviews.
Jsoup has a test suite with a relatively good instruction coverage of 86\%.
Our interviews focused on the classes \texttt{Attribute} and \texttt{AttributeTest}, as we expected the concept of an HTML attribute to be known by our participants.
\texttt{Attribute} is fairly well tested, with most functions covered by the test suite.
However, its custom implementation of \texttt{hashCode}, \texttt{clone} and several branches in \texttt{equals} were not covered.

To take the setup burden off of our participants, we set up an instance of IntelliJ with our plugin on a server and let the interviewees interact with it through the browser.
This was possible through the JetBrains Projector tool\footnote{\url{https://jetbrains.github.io/projector-client/mkdocs/latest/}}.

\subsection{Interview Procedure}
To get a rough context of the participant's opinion on and knowledge about software testing, we asked an open question about the participant's prior experiences with testing software.
Then we briefly introduced test amplification: automatically modifying existing test cases to generate new ones that improve the coverage and can be taken over into the test suite.
We explained that the goal of the interview is to see their interaction with our tool and gather feedback on what aspects they like, what they would change, and how they would use such a test amplification tool in their work.
We sent the developers a link to our online setup of IntelliJ which they accessed through their browser.
We introduced the example project and explained how to start \testcube.
From this point on we invited the interviewee to explore on their own, thinking aloud about all their thoughts and impressions.
We did not define an explicit task to solve, rather our introduction of test amplification and the user interface of \testcube animated the participants to browse through the test cases and judge whether to include them, and in some cases include them in the test suite of the example project.
We kept any more explanations to a minimum to observe a situation as close as possible to the developer interacting with the tool alone.
We let each participant amplify and browse through several test cases for about twenty minutes.
During this time we ask them to think aloud about their impressions.
We nudge them by asking questions about their actions and opinions on \testcube’s behavior.
At the end we asked them to fill out the \emph{System Usability Score} questionnaire, a metric frequently used in the field of Human-Computer Interaction to assess how useable a product is~\citep{bangor2008an-empirical}.
While filling out the questionnaire, we ask them to reflect on the usability of the plugin and how it could be adapted to better fit their needs.

\subsection{Data Collection and Analysis}
\label{sec:data-analysis}
We recorded all interviews, including the screen of the developer while they were interacting with \testcube.
In addition, the interviewer took extensive notes.
We performed open coding~\citep{corbin1990grounded} to analyze the interview notes, checking back with the recording when anything was unclear or missing from the notes.
Following that, we applied axial coding~\citep{corbin1990grounded} to structure the emerged codes.
We report our findings along these axial codes, which we assigned to each of the research questions.
\Cref{tab:results-overview} presents the axial codes arising from our analysis of the interviews.

All interviews and the initial coding were performed by the first author.
To increase the validity of our analysis, the second author watched two of the performed interviews, took notes and coded them separately.
Then we compared the codes both authors created for the validation interview and refined our coding schema and our interpretations of the interviews.
We saw that both focused on different aspects of the interviews, one assigning about 20 codes and the other one about 10 codes per interview.
In total, we agreed on 90\% of the assigned codes in the first validation step.
As a second validation step, we performed an inter-rater reliability analysis.
We selected re-occurring topics from our codes that appear in 4 or more interviews.
The second author assigned them to 3 further interviews.
To compare the assignments of both authors, we calculate the percentage agreement (70\%) and Cohen's Kappa (60\%, moderate agreement).
The value of Cohen's Kappa is relatively low, because some of the codes we validate have skewed values.
If a code appears in nearly all the cross-validated interviews, its chance of appearance is close to one, leading to a small Cohen's Kappa because arithmetically the agreement could be a coincidence.
We provide our code book together with the code's frequencies in the interviews, as well as our cross-validation ratings in our online appendix~\citep{brandt2021amplification-interview-replication}.

\section{Results}
\label{sec:results}
In this section, we present the results we elicited from our study.
Firstly, we give an overview of key demographic factors characterizing our study participants.
Secondly, we detail what factors are important to the developers when it comes to the generated test cases themselves (\RQ{1}) and which aspects make a test exploration tool developer-friendly (\RQ{2}).
Next, we describe the various kinds of information the developers sought while exploring the generated test cases (\RQ{3}) and what value our interviewees saw in automatic test generation (\RQ{4}).
Every \emph{observation} that comes from the interviews will be labeled with \observation[d]{n} and if it is directly tied to one of the codes we assigned, we also report its \emph{support}, i.e., in how many interviews we observed it. 
The observations without explicitly mentioned support summarize multiple codes, describe anecdotal evidence or report general impressions we obtained overarching the single interviews.
Even though we cannot link them to a specific code from our interview notes, we still consider them valuable to report for our qualitative study.

While high support signals that a topic is very relevant for our participants, we cannot infer from a small support number that an aspect is less relevant.
As we wanted to explore as many aspects of developer-centric test amplification as possible, we mainly let the comments of our participants guide the direction of the interviews, similar to an unstructured interview~\citep{zhang2009unstructured}.
This lead to many observations only appearing in a small number of interviews, possibly because the topic they concerned was only reached in a small number of interviews.

Throughout this section, we structure our explanation along the axial codes of our results presented in \Cref{tab:results-overview}.

\begin{table}[tbp]
\centering
\begin{tabular}{lll}
\toprule
 & & Identifiers 
\\
 & & Concise 
\\
 & \multirow{-3}{*}{Code} & Consistent 
\\ \cmidrule{2-3}
 & & Relevant 
\\
 & & Invariant 
\\
\multirow{-6.5}{*}{\makecell[c]{Generated Test Cases\\\RQ{1}}} & \multirow{-3}{*}{Behavior} & Diverging 
\\ \cmidrule{1-3}
 & & Minimal Configuration  
\\
 & & Integration 
\\
 & \multirow{-3}{*}{Ease of Use} & Usability 
\\ \cmidrule{2-3}
 & \multicolumn{2}{l}{\makecell{Information Management}}
\\
 & \multicolumn{2}{l}{Focus}
\\ \cmidrule{2-3}
 & & Runtime
\\
\multirow{-7}{*}{\makecell[c]{Exploration Tools\\\RQ{2}}} &\multirow{-2}{*}{\makecell{Expectation Management}} & Capabilities 
\\ \cmidrule{1-3}
 & & Behavior / Intent 
\\
 & & Outcome
\\
 & \multirow{-3}{*}{Test Case} & Runtime  
\\ \cmidrule{2-3}
 & \multicolumn{2}{l}{Code Under Test}
\\
 & \multicolumn{2}{l}{Coverage}
\\
\multirow{-6.5}{*}{\makecell[c]{Sought Information\\\RQ{3}}} & \multicolumn{2}{l}{Original Test Case} 
\\ \cmidrule{1-3}
 & & Ease Test Engineering 
\\
 & \multirow{-2}{*}{Improve Test Suite} & Inspiration 
\\ \cmidrule{2-3}
 & \multicolumn{2}{l}{Learning}
\\
\multirow{-4.5}{*}{\makecell[c]{Value for Developers\\\RQ{4}}} & \multicolumn{2}{l}{Confidence}
\\ \bottomrule
\end{tabular}%
\caption{Structured overview of the answers to our research questions. Shown are the axial codes we obtained during our data analysis described in \Cref{sec:data-analysis}.}
\label{tab:results-overview}
\end{table}

\begin{figure*}[tbp]

    \centering
        \begin{subfigure}[b]{0.45\textwidth}  
            \centering
            \includegraphics[height=2.5cm,
            page = 2,
            trim={6.3cm 17.7cm 8.5cm 5.5cm},
            clip
            ]{demographics-graphs.pdf}
            \caption[]%
            {{\small Years of Experience}}    
            \label{fig:experience-demographics}
        \end{subfigure}
        \hfill
        \begin{subfigure}[b]{0.45\textwidth}
            \centering 
            \includegraphics[height=2.5cm,
            page = 1,
            trim={3.8cm 19.5cm 11.5cm 4cm},
            clip
            ]{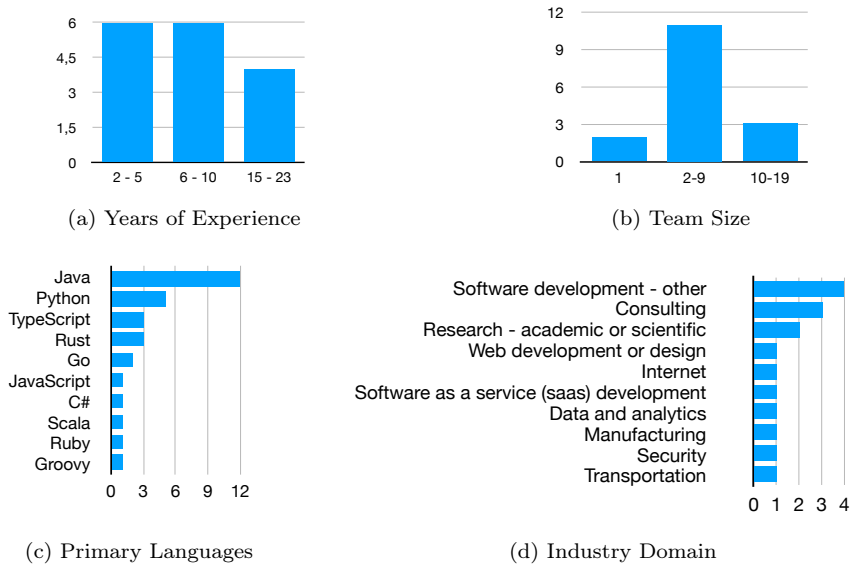}
            \caption[]%
            {{\small Team Size}}    
            \label{fig:teamsize-demographics}
        \end{subfigure}
        \vskip\baselineskip
        \begin{subfigure}[b]{0.35\textwidth}   
            \centering 
            \includegraphics[height=3.5cm,
            page = 4,
            trim={7cm 17cm 9cm 5.5cm},
            clip
            ]{demographics-graphs.pdf}
            \caption[]%
            {{\small Primary Languages}}    
            \label{fig:languages-demographics}
        \end{subfigure}
        \hfill
        \begin{subfigure}[b]{0.6\textwidth}   
            \centering 
            \includegraphics[height=3.5cm,
            page = 3,
            trim={5cm 17.5cm 7cm 5.5cm},
            clip
            ]{demographics-graphs.pdf}
            \caption[]%
            {{\small Industry Domain}}    
            \label{fig:industry-demographics}
        \end{subfigure}
        \caption
        {Summarized demographics of our interview participants.}
        \label{fig:demographics}
\end{figure*}

\subsection{Participants}
We recruited 16 participants for our study, whose demographics are summarized in \Cref{fig:demographics}.
As shown in \Cref{fig:experience-demographics}, their previous experience with software development was distributed in a range from two to 23 years.
Most of them work in teams of two to nine and consider Java to be among their primary programming languages.
Our participants work in a wide variety of industries, which are presented in \Cref{fig:industry-demographics}.

\subsection{RQ1: What Are the Key Factors to Make Amplified Test Cases Suited for Developer-Centric Test Amplification?}
\label{sec:rq1-results}

During the interviews, it quickly became clear to us how important the test cases themselves are for the developers.
Many of our participants were directly focusing on the test cases and spend much of their time praising or critiquing them \observation[d]{1}. 
This is reflected in the large number of observations with high support we present in this section.
What unified our participants is that they tried to understand the generated test cases \observation[d]{2}. 
Rojas et al. also noted that how easy it is to obtain the intent and behavior of a test case is a strong indicator for its quality~\citep{rojas2015automated}.

\paragraph{\textbf{Code: Identifiers}}
Looking at the \textbf{code} of the test cases, the most prevalent comment was the need for less cryptic \textbf{identifiers} \supobservation[d]{3}{14}.
The developers wished that the identifiers, such as the test name or variable names, convey the intent of the new test case \supobservation[d]{4}{2}.
They gave examples such as which code is additionally covered or simply which methods are newly called.
More expressive identifiers would help them understand the intent of the test case faster.
While integrating the test cases, many participants renamed the test cases \support{7} and the used variable identifiers \support{3}.
For identifier names that should be renamed by the developer during the inspection, two participants promoted a clearer naming such as ``\texttt{placeholderN}''.

\paragraph{\textbf{Code: Concise}}
We observed that the code should be as \textbf{concise} as possible.
Developers were confused by \textbf{unnecessary statements} \supobservation[d]{5}{7}, which were left over from the amplification process. 
Underneath them were object initializations or method calls no longer relevant for the intent of the new test case.
The developers needed additional time to detect these statements as unimportant and to delete them \supobservation[d]{6}{3}.

We saw a similar effect \textbf{with unnecessary casts} \supobservation[d]{7}{8} introduced by DSpot for type safety.
In many cases, the casts could be identified as superfluous, but the developers were unhappy that extra work is necessary to remove obviously superfluous code.

For generated assertions that check the return value of a function, DSpot actively splits the function call and the assertion through a variable declaration.
While one participant preferred this step for clarity and would find it easier to understand with an expressive variable name
, three other participants were annoyed by the additional bloating of the code \supobservation[d]{8}{3}.
While inlining the variable declaration a participant even pointed out that splitting the function call from the assert statement lead to \textbf{less powerful assertions} being used \observation[d]{9}.
Instead of \texttt{assertNotEquals} the test used \texttt{assertFalse}, which gives a much less expressive error message in case the assertion fails.
\texttt{assertNotEquals} throws a \texttt{org.junit.ComparisonFailure} and prints the expected and actual values, while \texttt{assertFalse} simply throws an \texttt{java.lang.AssertionError} at the line of the assertion.

Another issue brought up by the developers was \textbf{concise input strings}.
To generate parameters while creating new objects for the setup phase of a test case, DSpot can generate unnecessarily long random strings of characters \supobservation[d]{10}{2}.
To understand the test's behavior, developers now would need to know which of these characters are important for the intent of the test case, which takes a lot of time and effort.
In one case we observed, DSpot created a new object and checked that it is not equal to the existing one.
The developer spent a lot of time going through the various special characters in the constructor parameters for the new object, trying to determine which one is triggering the behavior of the test case.
In the end, none of the special characters were necessary, the strings simply had to be different from the strings initializing the existing object.
This anecdote points to the need for minimizing input values in generated test cases, which was also pointed out by Fraser et al.~\citep{fraser2013evosuite}

\paragraph{\textbf{Code: Consistent}}
Apart from the need for the test code to have a high quality~\cite{athanasiou2014test} itself, it should also be \textbf{consistent} with the rest of the test suite.
Three participants pointed out that the assertion methods were fully qualified instead of statically imported, like in the original test case \supobservation[d]{11}{3}.
A similar comment was made for the identifiers, which one participant wanted to be in the same naming schema as existing test cases.

\paragraph{\textbf{Behavior: Relevant}}
Moving to the behavior of the test cases, we saw that it is important to test methods \textbf{relevant} to the developers.
As our example project already had a relatively good test suite covering most core functions of the class, many of the proposed test cases covered extra branches in \texttt{equals}, \texttt{hashCode} or \texttt{clone}.
The initial reaction of various participants was ``I would not test this method'' \supobservation[d]{12}{11}, leading some to discard the test directly.
Others investigated further and uncovered that hashCode was overwritten with a custom implementation, which for one participant meant it was relevant after all to test the method.
We believe that it is not only important to focus on testing methods important to the developer, such as core functions defined in a class, but also to make it clear why a newly covered function is considered important, e.g., because it overwrites the defaults with a custom implementation.
How many interesting test cases are proposed was an important point for the developers we interviewed, this would majorly influence whether they keep using \testcube \supobservation[d]{13}{4}.

\paragraph{\textbf{Behavior: Invariant}}
A further comment on the behavior of the generated test cases was that developers would like them to test \textbf{invariants} of methods instead of absolute values \supobservation[d]{14}{8}.
The most prominent example being \texttt{hashCode}.
As other test generation tools, DSpot uses the current behavior of a system as an oracle.
To test \texttt{hashCode}, it calls the method on an object and creates an assertion comparing the resulting value to the return value of \texttt{hashCode}.
This is a fragile test case, as the \texttt{hashCode} changes as soon as any changes are made to the class's attributes.
Our participants advocated testing the invariant of \texttt{hashCode} instead of an absolute value.
Interestingly though, they proposed a variety of ways how to test this invariant: Cloning an object and checking the \texttt{hashCode} is still the same (as well as that they are equals), creating the same object twice and compare the \texttt{hashCode}, check that if equals is true the \texttt{hashCode} is also the same, or even creating random objects and verifying that only a few of them lead to hashCode collisions.
While proposing amplified test cases to open source projects, Danglot et al. also saw diverging reactions to test cases testing \texttt{hashCode}~\citep{danglot2019automatic}.

\paragraph{\textbf{Behavior: Diverging}}
One of our observations that is special to test amplification is how far the behavior of the generated test case should \textbf{diverge} from the original test case \supobservation[d]{15}{6}.
Some of our participants were enthusiastic that the generated test cases explored so many new paths and scenarios, even naming this as one of the key strengths of \testcube.
Others were confused by this divergence as they mainly focused on comparing the behavior of both test cases.
The most severe cases of such a divergence approach when the original test case involves objects from another class than the class under test.
Some of the amplified test cases then test functionality in this other class and completely disregard the original class under test \supobservation[d]{16}{3}.
While these tests can be valid and helpful additions to the test suite, our participants mostly disliked them, because they were focussing on testing the original class under test.
In a future version of \testcube, tests for another focal class should be marked as such and proposed to be added to the fitting test class.

\begin{rqanswerbox}{\textbf{RQ1: The Key Factors to Make Amplified Test Cases Suited for Developer-Centric Test Amplification}}
Summarizing the results and observations described in this section, the key factors to make amplified test cases suited for developer-centric test amplification are concerned with the code and the behavior of the test cases.
When it comes to code, the variable identifiers and test names should be meaningful, the code should be short and concise, and consistent in terms of quality and style with the rest of the test suite.
With respect to its behavior, an amplified test case should execute scenarios that are relevant for the developer, should test invariants in place of absolute values where possible, and should match the developer's expectation in terms of divergence from the original test case.
\end{rqanswerbox}

\subsection{RQ2: What Are the Key Factors to Make Test Exploration Tools Suited for Developer-Centric Test Amplification?}
\label{sec:rq2-results}

To enable developers to interact with and judge the amplified test cases, we created a test exploration tool.
In the following, we will explain our observations from the interviews on what factors are important for such a tool to be suited for developer-centric test amplification.

\paragraph{\textbf{Ease of Use: Minimal Configuration}}
First and foremost, a test exploration tool should be \textbf{easy to use} and especially \textbf{easy to start}.
Our approach of using a \textbf{default configuration} was successful, two of our participants pointed out how little effort was needed from their side to get started \supobservation[d]{17}{2}.
Some still noted concerns about how easy the tool would be to set up locally for their projects \supobservation[d]{18}{2}, so clear supporting documentation is important if one wants to let the developer try out and discover a tool all by themselves.

\paragraph{\textbf{Ease of Use: Integration}}
A factor that helped developers start up and explore \testcube so quickly was its tight \textbf{integration} with IntelliJ \supobservation[d]{19}{3}.
Participants noted that it was easy to start from the ``run test'' location, two made use of the built-in code navigation to explore the code under test \supobservation[d]{20}{2} and one liked that they could perform all actions without having to switch tools \supobservation[d]{21}{1}.

\paragraph{\textbf{Ease of Use: Usability}}
In addition to minimal configuration and being integrated, a developer-centric exploration tool should also adhere to the long-established criteria for \textbf{usability} from Human-Computer-Interaction research~\citep{bevan2001international}.
We have seen that it is important to give the developer control over the layout of \testcube: various participants had different wishes for which information they want to see and how much space the tool should take up on their screen \supobservation[d]{22}{1 each from 5 codes}.
Some were looking for buttons to close, e.g., the coverage editor they no longer needed \support{3} or got confused after they could not undo an unintentional action \support{3}.
We believe it is crucial for a successful tool to give the developer options to configure the layout of the tool to fit their needs and let them recover from errors.

With the help of the \emph{System Usability Score}, we evaluated the overall usability of \testcube.
44\% of our participants rated the usability as ``Excellent'', 38\% as ``Good'' and 19\% as ``Poor''.
This shows that even with the above mentioned issues, we overall succeeded in creating a tool that is easy to use.

\paragraph{\textbf{Information Management}}
We observed big difficulties with \textbf{managing the information} \testcube is displaying for developers.
The text detailing which instructions are additionally covered was overlooked by many study participants, some later said they thought it is  ``unimportant debug output'' \supobservation[d]{23}{3}.
Providing the information sought out by developers in a way that does not overwhelm them and is accessible to them where they expect it is one of the big challenges looking at future versions of \testcube.
Also the number of generated test cases should not be too large \supobservation[d]{24}{4}.
For some methods, over fifty new test cases were proposed that one by one tested a previously uncovered class.
One participant said they lost interest after looking over several of these test cases and seeing how many were left \supobservation[d]{25}{1}.
An effective test exploration tool should focus on a few impacting test cases to not overwhelm the developer and keep each interaction session compact.
Additionally it would help to rank the generated test cases and show the most impactful ones first.

\paragraph{\textbf{Focus}}
Related to information management is also the issue of \textbf{focus}.
Through nearly each one of our interviews, we saw how important it is to only show information to the developer which they are supposed to focus on in that moment.
In the current design of \testcube, the amplified test cases are all part of the same text file presented at once in an editor.
This means that multiple tests are visible at the same time.
While \testcube's internal navigation, e.g., the coverage information and the automatic adding to the test suite, was focused only on the first test case at the start, many of our participants started scrolling through the list of test cases immediately \supobservation[d]{26}{5}. 
Later some of them were confused \supobservation[d]{27}{3}, as they tried to add the test case they were currently focussing on to the test suite, while \testcube copied over the first one in the list.
It is therefore extremely important for a future test exploration tool to make sure the focus of the developer aligns with the focus of the tool.
For example, by only showing the code of one test at a time and therefore forcing the user to click on the next and previous buttons to explore the generated test cases.

\paragraph{\textbf{Manage Expectations: Runtime}}
A test exploration tool should \textbf{manage the expectations} of its users.
We observed this with the \textbf{runtime} of the amplification process.
Even though we took care in our configuration to keep the runtime of DSpot as low as possible, in some cases the generation still took several minutes to complete, which four participants considered as too long \supobservation[d]{28}{4}.
While we included a business indicator that signaled to the developers that the amplification is running a background task, many were wondering how long it will take before they get results.
Our participants wished for an expressive progress bar that either gives an estimation of the remaining time or at least shows an approximated form of progress \supobservation[d]{29}{2}.
Some were wondering whether, or even expecting that, it is possible to switch to another task while they were waiting.

\paragraph{\textbf{Manage Expectations: Capabilities}}
The expectations with respect to the \textbf{capabilities} of a tool should also be correctly set.
As we gave the developers only a minimal introduction to the tool, it was not clear for some whether the generated test cases are meant to replace the existing test case or are meant to be an addition to the test suite.
Toward the end of their interviews, participants pointed out that they slowly understand the power of the tool better and see clearer how they would employ it \observation[d]{30}.
One pointed out that the generated test cases were much more appreciated by him now that he understood the editing effort which was necessary before including them.

Overall, we observed a plethora of important aspects to make a test exploration tool developer-centric.
It should be easy to start and use, the way of displaying information needs to be carefully chosen, it has to keep track of the focus of the developer and manage the user's expectations towards its capabilities and runtime.

\begin{rqanswerbox}{\textbf{RQ2: The Key Factors to Make Test Exploration Tools Suited for Developer-Centric Test Amplification}}
In summary, a key factor to make test exploration tools suited for developer-centric test amplification is making the tool easy to use: through minimal configuration, through a tight integration into the developer's existing environment and through adhering to established usability principles.
Such tools should manage the information they present to the developer and help the developer focus on the information they need for their current task.
Further, a test exploration tool should manage the expectations their users have towards the runtime and the capabilities of the tool to ensure that these expectations can be fulfilled.
\end{rqanswerbox}



\subsection{RQ3: What Information Do Developers Seek While Exploring Amplified Test Cases?}
\label{sec:rq3-results}

While exploring the generated test cases, our participants did not only scrutinize the test code itself but were also looking for and asking about a lot of additional information.
We saw that it is crucial to provide quick and familiar ways for developers to provide this information so they can efficiently decide on whether to keep or how to adapt an amplified test case.

\paragraph{\textbf{Test Case: Behavior / Intent}}
As mentioned in \Cref{sec:rq1-results}, the test cases themselves and their \textbf{behavior} or rather their \textbf{intent} were a main focus of the developers. 
After making edits to a test case, one participant wondered whether the original intent of the generated case was still preserved \supobservation[d]{31}{1}. 
This is in line with Grano et al.'s results: developers are concerned with determining whether a unit test ``actually exercises the corresponding unit'' and how many relevant scenarios are covered~\citep{grano2020pizza}.
Prado and Vincenzi showed that the code of a test case is one of the main sources of information about a test case for the developer~\citep{prado2018towards}, an observation corroborated by Aniche et al.~\citep{aniche2021how-developers}.
As far as we observed, the current editor displaying the code of the generated test case is enough to satisfy this information need for developers.

\paragraph{\textbf{Test Case: Outcome}}
Furthermore, the developers were interested in the \textbf{outcome} generated test cases \supobservation[d]{32}{3}, i.e., whether they are passing or failing.
As all test cases generated by DSpot pass, this could be addressed by a better explanation of the tool.
Alternatively, tool developers could provide the existing IDE utility to run a test case in the editor proposing the new test case or provide functionality such as Infinitest, a tool that runs JUnit tests continuously in the background~\citep{infinitest2021infinitest}.
This would allow the developer to easily check that a test case is still passing after editing it and before integrating it to the test suite.

\paragraph{\textbf{Test Case: Runtime}}
The \textbf{runtime} of a test case was also pointed out by one of our participants \supobservation[d]{33}{1}, as they were used to projects where increasing the runtime of the continuous integration build was frowned upon.
Test exploration tools should include a note about the measured execution time with each test case.

\paragraph{\textbf{Code Under Test}}
While inspecting the new test cases, most of our participants quickly jumped to also inspecting the \textbf{code under test}.
Two were trying to understand its behavior \supobservation[d]{34}{2} to see the intent of the test case and to judge whether the additional coverage was relevant.
Also, they checked if the tested method overrides standard behavior and if exceptions were thrown and tested.
As Spadini et al. already pointed out, it is crucial for test review tools to provide easy navigation between test code and the code under test~\citep{spadini2018when}.
Prado and Vincenzi point out that developers should receive tool support to build the context between test code and code under test~\citep{prado2018towards}.
Through reusing the standard editor component of IntelliJ, \testcube allows its users to use their familiar code navigation tools, such as command-click to go to the definition of a method.

\paragraph{\textbf{Coverage}}
The third large area developers wondered about was \textbf{coverage}.
Under this falls the original coverage of the test suite and which additional coverage each generated test case and all the generated test cases together yield.
A recurring question was whether a functionality covered by the generated test cases was already covered by another test case \supobservation[d]{35}{2}.
Developers scanned the test suite to find other tests calling the same method.
Even though \testcube provides detailed information on which instructions are additionally covered by the new test case, not all participants understood that this implies the instructions were not covered by any existing test case.
The developer that found our visualization of the added coverage, found it helpful to see the covered lines not only as numbers but also in the code context \supobservation[d]{36}{1}.
They wished for a separate report of the original instruction coverage and the improvement of instruction coverage after including the amplified test cases.
We also observed two times that the developers used the coverage of a test case to infer its intent \supobservation[d]{37}{2}, an observation also made by Grano et al.~\citep{grano2020pizza}.
Sometimes it was unclear to our participants why the new code covers these additional instructions \supobservation[d]{38}{3}.
This points towards a need for exploration tools to visualize clearer how test code and code under test connect.

Instruction coverage seemed to be a satisfying metric for most of our participants.
Some of them wished for more information about how many branches are covered or hoped the tool would help them cover all branches as they would aim for while writing unit tests themselves.
Even though some of our participants were aware of the concept of mutation score, 
none of them asked for information about improved mutation score or for test cases that kill additional mutants.
Rojas et al. saw that in an industrial context many developers used coverage to evaluate a generated test case~\citep{rojas2015automated}.

\paragraph{\textbf{Original Test Case}}
Special for the case of test amplification, was the interest of the participants to inspect the \textbf{original test case} that was the basis for the amplification.
They tried to understand the intent of the original test case \supobservation[d]{39}{5} and used this information, together with the knowledge of which instructions were changed to determine the behavior of the amplified test cases \supobservation[d]{40}{2}.
Highlighting the changes from the original to the generated test case through comments \choice{4} was not successful in our study.
The developers ignored them and questioned their usefulness \observation[d]{41}.
We hypothesize that the generated test cases were short enough to spot the changes without the comments.

\begin{rqanswerbox}{\textbf{RQ3: The Information Developers Seek While Exploring Amplified Test Cases}}
In our interviews we observed that developers are interested in a wide array of information while exploring and inspecting amplified test cases.
For a test case itself, developers try to understand its behavior and intent, ask whether it is passing and how long it takes to execute.
Beyond the test case, they are concerned with the code under test, the original and added coverage as well as the original test case the amplification was based on.
\end{rqanswerbox}

\subsection{RQ4: What Value Does Developer-Centric Test Amplification Bring to Developers?}
\label{sec:rq4-results}

One way to make developer-centric test amplification successful, is to bring across the value they can expect from the amplified test cases and from using the test amplification tool.
To give us an indication, which values we should focus on, we collected comments from our participants about the benefits they believe they would achieve from using a tool similar to \testcube.

\paragraph{\textbf{Improve Test Suite: Ease Test Engineering}}
First and foremost, automatic test amplification would help them \textbf{improve their test suite}.
By proposing complete, ready-to-run test cases that cover more code or interesting behavior \supobservation[d]{42}{4}, automatic test amplification \textbf{eases test engineering} for developers.
The test amplification alleviates the developer from having to write test cases from scratch, reducing the effort necessary to develop a test suite.
Reducing effort is a concern for developers: one participant stated, that they would ``either look for less work or for tests with a better quality'' \supobservation[d]{46}{1}.

\paragraph{\textbf{Improve Test Suite: Inspiration}}
The generated test cases also provided \textbf{inspiration}.
Several users created new test cases to cover the behavior of the amplified test cases \supobservation[d]{47}{4}.
They were glad to be pointed to untested code paths \supobservation[d]{43}{4} and to unexpected scenarios that could happen in the system \supobservation[d]{44}{1}.
A recurring comment was that a test covers methods the participant always forgets to test \supobservation[d]{45}{3}.
By proposing new scenarios with the generated test cases, test amplification tools can take the burden of designing test scenarios of the developers.

\paragraph{\textbf{Learning}}
Packaging test case generation in an easily accessible plugin can be a valuable step to enable more developers to \textbf{learn} about test amplification itself.
Many of our participants did not know about the technique of test amplification before and one said that a plugin like \testcube could be a way to bring this idea into industry \observation[d]{48}.
The participants got more confident towards the end of the interviews about what \testcube can do for them and how they could apply it effectively \observation{30}.
In general, we saw that amplification was easy to grasp for the developers \observation[d]{49}.
A participant pointed out they would like to use such a tool while they are working on improving the test suite \supobservation[d]{50}{1} and another was eager to try it on their own projects \supobservation[d]{51}{1}.

\paragraph{\textbf{Confidence}}
One participant said that using \testcube more often would increase their \textbf{confidence} in their test suite \supobservation[d]{52}{1}.
On the one hand simply through the higher coverage after adding the generated test cases, and on the other hand because they see more important scenarios being covered.

\begin{rqanswerbox}{\textbf{RQ4: The Value Developer-Centric Test Amplification Brings to Developers}}
Our participants named a number of benefits they would gain from using an automatic test amplification tool regularly.
It would make it easier for them to develop test cases, by alleviating them from the effort to write the test cases and by providing inspiration of scenarios they tend to forget to test.
A developer-centric test amplification approach would support them learning about automatic test amplification and using it would increase their confidence in their test suite.
\end{rqanswerbox}


\section{Discussion and Recommendations}
\label{sec:discussion}

\begin{table}
\begin{tabular}{p{6.2cm}p{4.7cm}<{\raggedright\arraybackslash}}
Recommendation & Corresponding Observations
\\
\toprule
\textbf{Recommendation 1:} Consider the interaction of the developer with the test cases: provide a test exploration tool that is targeted towards the test generation method and integrated into the developer's environment.
& 
\observation{2} \observation{35} \observation{34} \observation{39} \observation{40} \observation{30}
\observation{43} \observation{31} \observation{24} \observation{26} \observation{27} \observation{28}
\\ \addlinespace
\textbf{Recommendation 2:} When the main goal is for developers to accept a test case into their maintained test suite, it is more important that the test case is understandable and relevant to the developer, than how much it impacts the coverage of the test suite.
& 
\observation{2} \observation{3} \observation{5} \observation{7} \observation{12} \observation{35} \observation{14}
\observation{11} \observation{9} \observation{10}
\\ \bottomrule
\end{tabular}%
\caption{Connection of our recommendations to our interview observations.}
\label{tab:recommendations-results}
\end{table}

In the following, we consolidate our results into two actionable recommendations on how to make amplified test cases and test exploration tools suited for developer-centric test amplification.
\Cref{tab:recommendations-results} shows from which of our interview observations we infer the recommendations.

We chose an additional layer between the test amplification and the developer, a test exploration tool, to address the issues users previously reported with DSpot.
Our prototypes could already surface different kinds of information the developers were seeking, such as the behavior of the test case \observation{2}, its coverage \observation{35}, or which code is tested \observation{34}.
Further, it became clear how tightly the characteristics of the test exploration tool are bound to the kind of test cases it presents to the developer.
In our design, the technique of test amplification \choice{1} 
and the information from the amplification process reports \choice{9}, is tightly bound to what our exploration tool \testcube presents to its users about the test cases \choice{14}.
Our participants were questioning how the test cases are generated, and also sought information especially related to test amplification, like the original test case \observation{39} \observation{40}.
We saw the importance of expectation management \observation{30}, e.g., on how much they should edit the proposed test cases, and conveying the value the test amplification can bring to the developer, such as pointing to untested code paths \observation{43}.
The tight integration into the developer's IDE was helpful to get them started quickly \observation{19}.
Overall, we saw a positive effect of using a test exploration tool to facilitate the developer-centric test amplification.
We conjecture that this support of an integrated test exploration tool is also beneficial for other test generation approaches that aim to be developer-centric.
We recommend to future authors of developer-centric test generation approaches to provide a test exploration tool that is targeted towards the test generation method they employ and accessible to the developer from their familiar environment.

\begin{intentionbox}{\textbf{Recommendation 1}}
  Consider the interaction of the developer with the test cases: provide a test exploration tool that is targeted towards the test generation method and integrated into the developer's environment.
\end{intentionbox}

Concretely, \testcube can be improved in several points: Clearer visualization of the connection from the amplified test case to the additionally covered instructions in the code under test \observation{23} \observation{34} \observation{38} and describing the behavior of the amplified test case and how it diverges from the original test case \observation{31}.
Further we can help developer focus by proposing one test case at a time \observation{24} \observation{26} \observation{27} and address waiting time \observation{28} by generating test cases before they are requested.

Looking at the results of our first research question, we can see that the developers were mainly concerned with \emph{understanding} the test cases \testcube presented to them \observation{2}.
The observations which occurred in most interviews are about the identifiers \observation{3}, the conciseness of the code \observation{5} \observation{7}, and trying to understand the behavior and intent of the test case \observation{31}, be it through the test code itself or the various other kinds of information sought.
During our interviews, the understanding was always the first step---only after the participants understood a test case they started to judge the impact or relevance \observation{12} \observation{35}.
When judging the test cases, we observed that not all tests which increase instruction coverage are relevant to developers, e.g., because they test a to them less important method \observation{12} or a too narrow behavior \observation{14}.
From these results, we infer that for a developer-centric approach, where the central aim is for a developer to take over the generated test case into their maintained test suite, the understandability of the generated test case and the relevance to the developer is of a bigger concern than how high its numeric impact is on the coverage of the test suite.
An understandable test case with a weaker coverage contribution is more likely to be accepted by developers, compared to a test case that increases coverage greatly but they discard because they can not understand what it does.
We recommend to future authors of developer-centric test generation approaches to prioritize the understandability of the generated test cases and their relevance to the developer higher than their impact on the coverage of the test suite.

\begin{intentionbox}{\textbf{Recommendation 2}}
  When the main goal is for developers to accept a test case into their maintained test suite, it is more important that the test case is understandable and relevant to the developer, than how much it impacts the coverage of the test suite.
\end{intentionbox}

Concretely, the amplified test cases generated by DSpot can be improved by generating useful identifiers \observation{3}, possibly informing about the unique coverage provided by the test case \observation{35}.
Further unnecessary statements \observation{5} and casts \observation{7} should be removed, the style of the test cases \observation{11} can be adapted to fit the existing test suite, randomly generated strings shortened and focused to the part triggering the tested behavior \observation{10}, and assertions adapted to use the most specific assertion giving an informative error message \observation{9}.

Our participants pointed out how easy it was to interact with \testcube right from their IDE \observation{17} \observation{19} \observation{20}, many found test cases that they liked and added them into the test suite of our example project.
We conjecture that combining the already powerful state-of-the-art test amplification approaches with well-designed, developer-centric test exploration tools will let us reach more developers to amplify their software testing practice.

\section{Threats to Validity}
There are several threats to the validity of our results which we discuss in this section.

\paragraph{Confirmability}
To ensure that our results are formed by the interviewees and not by the authors, we base our results as closely as possible on the interviews.
While coding and analyzing the interviews we performed extra steps to validate the codes elicited from the interviews and evaluated the inter-rater reliability, as described in \Cref{sec:data-analysis}.
Nevertheless, other researchers might structure the resulting codes differently or draw varying conclusions from them.
We publish the full codes together with their frequency in our interviews~\citep{brandt2021amplification-interview-replication} for others to further explore the research area and add to our study.

\paragraph{Reactivity and Respondent Bias}
As the first author created the prototypes and conducted all interviews, the statements of the participants might be influenced by the participants wanting to please the creator of the tool they are evaluating.
To mitigate this threat, we repeatedly invited the participants to be critical and refrained from defending the current state of the tool.
Based on the wide variety of critical and positive points we could collect, we conjecture to have mitigated this threat.

\paragraph{Construct Validity}
A threat to the construct validity of our study is that our participants interacted with an early prototype showing one possible design of a test amplification tool.
Bugs in the prototype or design decisions we took could influence the developer's experience and the generalizability of the results to general test amplification approaches.
We identify several of our results as being related to our choice of test amplification to generate the test cases \choice{1}, which we indicated while reporting them in \Cref{sec:results}.
Similarly, our observations can be influenced by our default configuration of DSpot.
Optimizing the configuration of DSpot to fit the target project would likely lead to more relevant test methods being generated.
Furthermore, our participants were not developers of the example project we used in the study.
We expect that developers familiar with a project would spend less time on understanding the original and amplified test cases and could judge more easily if a production method is relevant to be tested.

\paragraph{Dependability}
Whether our results are consistent and can be repeated in a replication is the concern of dependability.
With 16 participants we were able to interview a relatively large number of software developers.
Our presented results mainly focus on observations that we made in multiple interviews (that have high support).
Nevertheless, there were many insightful comments that only emerged from one or a few interviews.
Through the openness of our setup and questions, the interviews went in many different directions and the observations we could make are dependent on the taken direction.
We expect that repeating this study would yield different support for the rarely-mentioned aspects, however the 
overall conclusions will likely stay the same.

\paragraph{External validity}
There are several threats to the generalizability of our results.
As well as other state-of-the-art test generation tools, our prototypes address Java and its specific properties.
We expect our results to generalize to other object-oriented, statically typed languages and are curious to see the different information needs developers of other programming languages have.

The choice of presenting our prototypes together with the example project Jsoup can also impact our results.
Because equals, hashCode and clone were not covered by the existing test suite, DSpot generated tests mainly for these functions that were named ``irrelevant'' by several of our participants.
In other projects whose test suite has a lower or differently distributed coverage, the aspect of testing relevant methods might be less apparent.

As we performed convenience sampling, the results of our study might be influenced by our professional networks, as well as a self-selection bias of developers that are especially interested in high-quality test suites.
From the demographic information we collected, we conclude that we sampled from a broad variety of experiences, industry domains and team sizes.

\section{Related Work}
Various past works have focused on the two main parts of or approach, mainly improving the understandability of test cases and integrating test generation tools into development environments.

\subsection{Understandability of Test Cases}
The issues of cryptic identifiers and lack of documentation in generated test cases are addressed by Roy et al.~\citep{roy2020deeptc} in their tool \emph{DeepTC-Enhancer}.
With a combination of templates and deep learning, they generate comments that explain the behavior of a test case and meaningful identifiers.
Their work is an extension of \emph{TestDescriber} by Panichella et al.~\citep{panichella2016the-impact} and was evaluated by 36 developers.
The developers were most enthusiastic about the meaningful identifiers, while some said the explanatory comments are not concise enough.
In our interviews, we also observed the importance of expressive test names and variable identifiers.
While our participants were trying to understand the behavior of the amplified test cases, they could interpret the raw code of the test cases well.
Therefore, we do not believe any additional summarization of the test itself is necessary.
Easier access to the code under test and information about previous and added coverage are more relevant concerns going forward.
In similar vein, Li et al. describe UnitTestScribe~\citep{li2016automatically}.

Alsharif et al.~\citep{alsharif2019what} investigated which factors are important for the understandability of automatically generated SQL schema tests.
They saw that human-readable string values are better to understand than randomly generated ones and the repetition between generated test cases made it easier to focus on the relevant differences of the test cases towards each other.
Their results align with ours: Randomly generated strings were mentioned as confusing and our interview participants repeatedly used the similarity between the original and the amplified test case to understand the behavior and impact of the newly generated test case.

Daka et al.~\citep{daka2015modeling} define a regression model for test case readability based on various syntactic properties of test cases.
They integrate the model into the fitness function of EvoSuite~\citep{fraser2011evosuite} to generate more readable test cases.
In their model and post-experiment survey they identified several important factors overlapping with our findings:
Identifiers are important for the understandability of a test case, as well as no unnecessarily defined variables and short string literals.

Next to \emph{DeepTC-Enhancer} by Roy et al.~\citep{roy2020deeptc}, several further works focus on generating meaningful names for test cases.
\emph{NameAssist} by Zhang et al.~\citep{zhang2016towards} infers test names from the class under test, the expected outcome stated in the assertion and the overall test scenario defined in the body of the test.
Daka et al.~\citep{daka2017generating} derive test names from additionally covered exceptions, methods, outputs and inputs of the component under test.
They showed that the generated names are equally excepted compared to names given by developers and made it easier for developers to match a test to the code under test.
Including an advanced name generation approach such as the one by Daka et al.~\citep{daka2017generating} would be a valuable addition to \testcube and DSpot.

Bihel and Baudry~\citep{bihel2018adapting} focused specifically on making tests amplified by DSpot more accessible for developers.
They generate a natural language description of the changes made during the amplification, of the value observations which lead to new assertions, and of the mutants which will be killed by the newly added test cases.
These descriptions are designed to accommodate a pull request proposing to add an amplified test case.
In comparison, \testcube focuses on a just-in-time interaction of the developer, embedding test amplification into their IDE.
In our scenarios, not only tool performance, but also the amount of presented information is a distinguishing challenge.
Compared to Bihel and Baudry's approach~\citep{bihel2018adapting}, \testcube more carefully selects the information presented to the developer.
We also evaluate our approach in a study with developers.

Because of the high computational cost of test generation, many tools have opted for integration into the continuous integration process~\citep{arcuri2016unit,danglot2020an-approach}.
This, however, leads to a long time distance between triggering the test generation and receiving results~\cite{beller2017oops}, as well as the developers having to inspect the tools outside of their familiar development environment.
To provide more immediate value and direct feedback, we opted to let \testcube run on our user's computers, leading to many more constraints regarding the available execution power and therefore possible complexity of the applied algorithms.

\subsection{Test Generation Tools Integrated in the IDE}
Several other test generation tools have been integrated into IDEs up until now.
Following an industrial study of EvoSuite, Rojas et al.~\citep{rojas2015automated} pointed out the importance to integrate test generation tools into development environments.
Since then, EvoSuite has been lightly integrated into IntelliJ IDEA as a plugin~\citep{arcuri2016unit} which provides options to configure the test generation within an existing build process.
In contrast to this, \testcube runs independently from a project's build process and can be installed and applied with nearly no configuration\footnote{In the current version the user only has to provide the path to their Java 8 installation and their Maven Home.}.

DSpot has been integrated into the Eclipse IDE as a plugin together with other tools from the STAMP project~\citep{stamp2019ide}.
The plugin offers a graphical interface to set the various configuration parameters of DSpot and start the amplification process.
Compared to \testcube , the additional information showing the impact of a generated test case is just presented as a JSON text and the developer is still confronted with many configuration parameters.

Tillmann and de Halleux~\citep{tillmann2008pex-white} developed the Pex tool which generates inputs for parameterized tests based on program analysis.
They integrated their tool into Visual Studio, enabling the developer to generate and execute the unit tests by right-clicking on the parameterized unit test.
The tool presents the generated inputs and corresponding test results in a new window as a simple table.

\subsection{Interactive Test Generation}
The idea of connecting the developer closer with the test generation is also realized in \emph{Interactive Search-Based Software Testing (ISBST)}.
In the concept of Marculescu et al.~\citep{marculescu2012a-concept}, domain experts decide the importance of different components in the fitness function leading the automatic optimization of the test cases.
The interaction happens during the search process, where in defined moments the expert evaluates the current candidate test cases and adapts the fitness function for the next round of test generation.
When compared to manual testing, ISBST could find different test cases and execute behavior previously not considered by developers, similar to what our participants reported about \testcube \observation{45}.
Marculescu et al. also investigated the mental workload of developers using ISBST compared to manually writing test cases.
They did see a higher load and explain it through the distance from the developer's interaction with the fitness function to the outcome of the search process.
Similarly, we saw during our interviews that the developers tried to retrace the generation of the test cases, strengthening the choice to perform only small edits during the amplification to make the process easier to retrace.
After transferring their approach to industry~\citep{marculescu2018transferring}, Marculescu et al. point out the need for ISBST and other automated test systems to effectively communicate their results to their users.
Our work addresses this by prototyping a developer-centric test exploration tool and eliciting the key factors to make such tools suited to be used for test amplification.


\section{Conclusion}
With this paper, we are setting a step towards test amplification that is centered around the developer and their needs.
Based on reported issues with current state-of-the-art tools, we devised design intentions for a developer-centric test amplification approach that aims to generate test cases that will be taken over into the manually maintained test suite.
We used these intentions to adapt DSpot‘s test amplification and create \testcube, a powerful test exploration plugin for IntelliJ.
With the help of these tools, we interviewed 16 software developers from a variety of backgrounds and collected detailed insights on how the amplified test cases and the exploration tool should be adapted to best fit their needs.
Through evaluating the information sought during the test exploration, as well as the value test amplification brings to developers, we guide future tool developers on what they should bring forward in their upcoming, developer-centric test generation tools.
We summarized our observations and results into two recommendations:
Tool makers should consider the interaction of the developers with the amplified test cases and provide a targeted and integrated test exploration tool.
If taking over the test cases into the maintained test suite is the declared goal, the understandability of the amplified test cases should be prioritized over optimizing the coverage of the test suite.

In short, we contribute:
\begin{itemize}
    \item two recommendations on how to design developer-centric test amplification tools
    \item a structured overview of the key factors to make amplified tests as well as test exploration tools suited for developer-centric test amplification
    \item a refined, developer-centric test amplification approach, based on the DSpot test amplification
    \item a developer-oriented test exploration plugin for the IntelliJ IDE
\end{itemize}

Going forward we want to understand the different aspects of developer-centric test amplification in more depth.
We want to look into generating meaningful identifiers fast enough, ranking test cases according to their relevance to the developer, and providing them information such as runtime or coverage when they look for it, but without overwhelming them.
Our tools will dive deeper into their day-to-day development, for example by helping them incrementally generate test cases for new or untested classes.
We want to give them more power to direct the amplification and receive test cases that cover code or scenarios they are interested in, while also providing them with subtle, helpful recommendations before they realize they need another test case.
Our vision is to build tools and methods that empower developers to create better test suites with less effort, while they are at the steering wheel deciding over, leading, and benefiting from our automatic test amplification.


\begin{acknowledgements}
We would like to thank the participants of our study for the valuable feedback on our work. This work was sponsored by the Dutch science foundation NWO through the Vici ``TestShift'' project (No. VI.C.182.032).
\end{acknowledgements}

%
%

\bibliographystyle{spbasic}      
\bibliography{paper}   

\end{document}